\documentclass[5p]{elsarticle}
\usepackage{lmodern}
\usepackage[utf8]{inputenc}
\usepackage[T1]{fontenc}
\usepackage[tbtags]{mathtools}
\usepackage{mleftright}\mleftright
\usepackage{MnSymbol}
\usepackage{dsfont}
\usepackage{slashed}
\usepackage[squaren,Gray]{SIunits}
\usepackage{booktabs}
\usepackage{widetable}
\biboptions{square,comma,numbers,sort&compress}
\usepackage{hyperref}

\newcommand{\overbar}[1]{\mkern 1.5mu\overline{\mkern-1.5mu#1\mkern-1.5mu}\mkern 1.5mu}
\newcommand{\MSbar}{\overbar{\text{MS}}}
\newcommand{\SU}[1]{\ensuremath{\text{SU}(#1)}}
\newcommand{\lvec}[1]{\reflectbox{$\!\vec{\,\reflectbox{$#1$}}$}}
\newcommand{\lrvec}[1]{\mathrlap{\reflectbox{$\!\vec{\,\reflectbox{$#1$}}$}}\,\,\vec{\!\!#1}}
\newcommand{\vect}[1]{\ensuremath{\boldsymbol{#1}}}
\renewcommand{\overleftarrow}[1]{\lvec{#1}}
\renewcommand{\overrightarrow}[1]{\vec{#1}}
\renewcommand{\overleftrightarrow}[1]{\lrvec{#1}}
\newcommand{\ihat}{\skew{1}{\hat}{\vect{\imath}}}
\newcommand{\jhat}{\skew{4}{\hat}{\vect{\jmath}}}
\DeclareTextSymbolDefault{\textperiodcentered}{OMS}
\renewcommand{\cdot}{\mathbin{\mbox{\textperiodcentered}}}

\let\oldding\ding
\renewcommand{\ding}[1]{\ifnum#1=73 $\dagger$\else\oldding{#1}\fi}
\begin{document}
\begin{frontmatter}
\author[cor1,cor2]{G.S.~Bali}
\author[cor1]{V.M.~Braun}
\author[cor1]{M.~{G\"ockeler}}
\author[cor1]{M.~{Gruber}}
\author[cor1]{F.~{Hutzler}}
\ead{fabian.hutzler@ur.de}
\author[cor1,cor3]{P.~{Korcyl}}
\author[cor1]{B.~{Lang}}
\author[cor1]{A.~{Sch\"afer}}
\address[cor1]{Institut f\"ur Theoretische Physik, Universit\"at Regensburg, 93040 Regensburg, Germany}
\address[cor2]{Department of Theoretical Physics, Tata Institute of Fundamental Research, Homi Bhabha Road, Mumbai 400005, India}
\address[cor3]{Marian Smoluchowski Institute of Physics, Jagiellonian University, ul.\ \L ojasiewicza 11, 30-348 Krak\'ow, Poland}
\title{Second moment of the pion distribution amplitude\\with the momentum smearing technique\tnoteref{t1}}
\tnotetext[t1]{RQCD Collaboration}
\begin{abstract}
Using the second moment of the pion distribution amplitude as an example, we investigate whether lattice calculations of matrix elements of local operators involving covariant derivatives may benefit from the recently proposed momentum smearing technique for hadronic interpolators. Comparing the momentum smearing technique to the traditional Wuppertal smearing we find---at equal computational cost---a considerable reduction of the statistical errors. The present investigation was carried out using $N_f=2+1$ dynamical non-perturbatively order~$a$ improved Wilson fermions on lattices of different volumes and pion masses down to $\unit{220}{\mega\electronvolt}$.
\end{abstract}
\end{frontmatter}
\section{Introduction}
Many quantities of interest in high-energy physics involve 
hadrons carrying large momenta. The prime example is provided by form factors,
but also parton distribution functions (PDFs) and their generalizations,
in particular transverse momentum dependent parton distribution functions
(TMDs) receive their physical interpretation in the large-momentum limit.

Very high accuracy is expected for future experimental data,
e.g., on hard exclusive and semi-inclusive reactions at the JLAB $\unit{12}{\giga\electronvolt}$ upgrade~\cite{Dudek:2012vr} and at the 
Electron Ion Collider (EIC)~\cite{Boer:2011fh}, as well as on
$B$-meson decay and pion transition form factors at Belle~II at
KEK~\cite{Abe:2010gxa}. This accuracy has to be matched by an
increased theoretical precision. Such processes are usually studied
using factorization techniques, where the nonperturbative input is reduced
to operator matrix elements which, ideally, should be computed using
lattice QCD. Also in the cases where no momentum transfer
takes place between the initial and the final state one usually needs
to realize hadron sources with nonvanishing momenta in order to have the
possibility to employ operators with sufficiently simple renormalization patterns. 
It has also been argued~\cite{Braun:2007wv,Ji:2013dva} that hadron sources with large momenta offer novel opportunities, enabling a more direct calculation of parton distributions and hadronic light-cone wave functions by performing a collinear factorization of suitably chosen Euclidean correlation functions (e.g., ``quasi-PDFs''~\cite{Ji:2013dva}), thereby circumventing the traditional Wilsonian local operator product expansion.

Although the problem is known for quite some time, 
up to very recently~\cite{Bali:2016lva} no satisfactory techniques for
hadrons carrying high momenta on the lattice existed to suppress excited
state contributions while maintaining acceptable signal-to-noise ratios.
The generic method of reducing excited state overlaps consists of employing
carefully tuned extended interpolators. However, for larger momenta the usual
smearing techniques become increasingly less effective.
The basic idea of Ref.~\cite{Bali:2016lva} was to modify the usual quark
smearing functions by additional phase factors such that the centre of the
distribution in momentum space is shifted towards the desired value. By
implication, such smearing functions correspond to oscillating wave packets
in position space. 

It was shown that this technique, which we will refer to as momentum smearing,
leads to considerably improved signal-to-noise ratios
for the pion and the nucleon two-point functions~\cite{Bali:2016lva} as well
as for lattice
observables that are related to quasi-PDFs~\cite{Alexandrou:2016eyt}.
In this letter we address another class of applications, namely
computing hadronic matrix elements that contain local operators with
covariant derivatives, e.g., moments of parton distributions
and distribution amplitudes (DAs)~\cite{Martinelli:1987si,Braun:2006dg,Arthur:2010xf,Braun:2014wpa,Braun:2015axa,Bali:2015ykx}.
In the case that we specifically study, i.e., moments of DAs,
the matrix elements of interest are proportional to powers of the
hadron momentum and are known, empirically, to be very noisy when
using traditional methods. 
We will demonstrate that momentum smearing results in a major
improvement of the quality of the signal for the second moment of the pion DA.
In fact, it turns out that this technique is so effective that, at small
pion masses and large lattice volumes, statistical fluctuations can be
further reduced by deliberately selecting a momentum that is larger than
the smallest possible~choice.

The scope of the present study is mainly methodological. In addition,
we present the first lattice calculation of the 2nd moment of the pion DA
using $N_f=2+1$ dynamical clover Wilson fermions.
The results are compatible with the latest $N_f=2$ study~\cite{Braun:2015axa},
while the second moment is somewhat smaller than what has been reported
in a simulation employing $N_f=2+1$ domain wall fermions, which has been carried
out at a coarser lattice spacing and at larger quark masses~\cite{Arthur:2010xf}. 
\section{General formalism}
\subsection{Continuum definitions}
Pseudoscalar mesons like the pion have only one independent leading twist (twist two) DA, $\phi$, which is defined via a meson-to-vacuum matrix element of renormalized non-local quark-antiquark light-ray operators,
\begin{align} \label{Bethe}
\begin{split}
\MoveEqLeft \langle 0 | \bar{d}(z_2 n) \slashed{n} [z_2n,z_1n] \gamma _5 u(z_1 n) | \pi^+ (p) \rangle =  \\
&= i f_\pi (p\cdot n) \int _0 ^1\! dx\, e^{-i(z_1x+z_2(1-x))p\cdot n} \phi  (x,\mu^2),
\end{split}
\end{align}
where $z_{1,2}$ are real numbers, $n^\mu$ is an auxiliary light-like vector with $n^2=0$, and $|\pi^+(p)\rangle$ represents the ground state pseudoscalar $\pi^+$ meson with on-shell momentum $p^2=m_\pi^2$. The straight path-ordered Wilson line connecting the quark fields, $[z_2n,z_1n]$, is inserted to ensure gauge invariance. The scale dependence of $\phi$ is indicated by the argument $\mu ^2$.%
\par%
Neglecting both isospin breaking and electromagnetic effects, the DAs of the charged pseudoscalar $\pi ^\pm$ and the neutral $\pi ^0$ are trivially related such that it is sufficient to consider only one of them.
The decay constant $f_\pi$ appearing in Eq.~\eqref{Bethe} can be obtained
as the matrix element of a local operator,
\begin{align}
\langle 0 | \bar{d}(0)\gamma _0 \gamma _5 u(0) | \pi^+ (p) \rangle = if_\pi p_0,
\end{align}
and has a value of $f_\pi\approx\unit{130}{\mega\electronvolt}$~\cite{Rosner:2015wva}.%
\par%
The physical interpretation of Eq.~\eqref{Bethe} is that the fraction~$x$ of the pion momentum is carried by the $u$~quark, while the $\bar{d}$~antiquark carries the remaining fraction~$1-x$. Hence the difference of the momentum fractions,
\begin{align}
 \xi = x-(1-x) = 2x-1,
\end{align}
contains all nontrivial information and its moments are defined as
\begin{align}
\label{Moments}
 \langle \xi ^n \rangle = \int _0 ^1\!\! dx\, (2x-1)^n \phi (x,\mu^2).
\end{align}
\par%
Since the Gegenbauer polynomials $C_n ^{3/2} (2x-1)$, which correspond to irreducible representations of the collinear conformal group $\text{SL}(2,\mathds{R})$, form a complete set of functions, the DAs can be expanded as
\begin{align}
 \phi (x,\mu ^2) = 6x(1-x) \left[ 1+ \smashoperator{\sum _{n=1} ^\infty } a_n (\mu ^2) C^{3/2}_n (2x-1) \right],
\end{align}
where the Gegenbauer moments $a_n$ renormalize multiplicatively in leading logarithmic order. Higher-order contributions in the Gegenbauer expansion are suppressed at large scales, since the anomalous dimensions of~$a_n$ increase with~$n$. Hence, in the asymptotic limit $\mu\rightarrow\infty$ only the leading term survives, which gives:
\begin{align}
\phi  (x,\mu \rightarrow \infty ) = \phi  ^{\rm{as}}(x)=6x(1-x).
\end{align}
\subsection{Lattice definitions}
From now on we will work in Euclidean spacetime and follow the conventions of Ref.~\cite{Braun:2015axa}. The renormalized light-ray operator on the left-hand side of Eq.~\eqref{Bethe} generates renormalized local operators. This means that Mellin moments of the DAs, see Eq.~\eqref{Moments}, can be expressed in terms of matrix elements of local operators and can be evaluated using lattice QCD. In order to calculate the second moment of the pion DA ($n=2$), we define the bare operators
\begin{align}
 \mathcal{P}  (x) &= \bar{d}(x) \gamma _5 u(x), \\
 \mathcal{A} _\rho (x) &= \bar{d}(x)\gamma _\rho \gamma _5 u(x), \\
 \mathcal{O}^- _{\rho\mu\nu} (x) &= \bar{d}(x) \left[ \overleftarrow{D}_{(\mu} \overleftarrow{D}_{\mathstrut\nu} -2 \overleftarrow{D}_{(\mu} \overrightarrow{D}_{\mathstrut\nu} + \overrightarrow{D}_{(\mu} \overrightarrow{D}_{\mathstrut\nu} \right] \gamma _{\rho )} \gamma _5 u(x), \\
 \mathcal{O}^+ _{\rho\mu\nu} (x) &= \bar{d}(x) \left[ \overleftarrow{D}_{(\mu} \overleftarrow{D}_{\mathstrut\nu} +2 \overleftarrow{D}_{(\mu} \overrightarrow{D}_{\mathstrut\nu} + \overrightarrow{D}_{(\mu} \overrightarrow{D}_{\mathstrut\nu} \right] \gamma _{\rho )} \gamma _5 u(x),
\end{align}
where $D_\mu$ is the covariant derivative, which will be replaced by a symmetric discretized version on the lattice. In order to obtain a leading twist projection we symmetrize over all Lorentz indices and subtract all traces. This is indicated by $(\dots)$, for example $\mathcal{O}_{(\mu \nu)} = \frac{1}{2} \left( \mathcal{O}_{\mu \nu}+\mathcal{O}_{\nu \mu} \right)-\frac{1}{4}\delta _{\mu \nu} \mathcal{O}_{\lambda \lambda}$.  By using the shorthand notation $\overleftrightarrow{D}_\mu = \overrightarrow{D}_\mu - \overleftarrow{D} _\mu$, the operator $\mathcal{O}^- _{\rho\mu\nu}$ can also be written as
\begin{align}
 \mathcal{O}^- _{\rho\mu\nu} (x) &= \bar{d}(x)  \overleftrightarrow{D}_{(\mu} \overleftrightarrow{D}_{\mathstrut\nu}  \gamma _{\rho )} \gamma _5 u(x).
\end{align}
The operator $\mathcal{O}^+ _{\rho\mu\nu}$ is, in the continuum, given by the second derivative of the axial-vector current:
\begin{align}
 \mathcal{O}^+ _{\rho\mu\nu} (x) = \partial _{( \mu} \partial _{\mathstrut\nu} \mathcal{A} _{\rho )} (x).
\end{align}
This is not the case on the lattice due to discretization effects of the derivatives which can be numerically sizable. The mixing with operators of lower dimension can be prevented by selecting lattice operators that belong to a suitable irreducible representation of the hypercubic group $\mathrm{H}(4)$~\cite{Arthur:2010xf,Braun:2006dg}. For our case, this corresponds to choosing all indices different for the operators $\mathcal{O}^\pm$. Identifying one index with the temporal direction, this leaves us with the operators
\begin{align}
 \mathcal{O}_{4jk} ^\pm , \quad j\neq k.
\end{align}

In order to extract the desired moments we use two-point correlation functions of the operators $\mathcal{O}_{4jk} ^\pm$ and $\mathcal{A}_\rho$ with an interpolating field,
\begin{align}
C_\rho (t,\vect{p}) &= a^3 \sum _{\vect{x}} e^{-i\vect{p}\vect{x}} \langle \mathcal{A}_\rho (\vect{x},t)J^\dagger(0) \rangle , \\
C_{\rho \mu \nu}^\pm (t,\vect{p}) &= a^3 \sum _{\vect{x}} e^{-i\vect{p}\vect{x}} \langle \mathcal{O}_{\rho \mu \nu}^\pm (\vect{x},t)J^\dagger(0) \rangle,
\end{align}
where $J=\mathcal{P}$ or  $J=\mathcal{A}_4$. For sufficiently large~$t$, the ground state dominates and the correlation functions give
\begin{align}
\begin{split}
 C _{\mathcal{O}} (t,\vect{p}) &= \frac{1}{2E} \langle 0 | \mathcal{O}(0) | \pi^+(p) \rangle \langle \pi^+(p) | J^\dagger(0) | 0 \rangle \\
 &\quad \times \bigl( e^{-Et}  + \tau _\mathcal{O} \tau _J e^{-E(T-t)} \bigr),
\end{split}
\end{align}
where the sign factors $\tau_\mathcal{O},\tau_J = \pm 1$ depend on the transformation properties of the correlation functions under time reversal.
\par
Following Ref.~\cite{Braun:2015axa}, the required matrix elements for the second moments can be extracted from the ratios
\begin{align}
\label{ratio}
\mathcal{R}^\pm _{4 i j} &= \frac{C^\pm _{4 i j}(t,\vect{p})}{C_4 (t,\vect{p})} = -p_i p_j R^\pm,
\end{align}
where $\langle \xi ^2 \rangle ^{\rm{bare}} = R^-$ and $a^{\rm{bare}} _2 = \frac{7}{12}(5R^- - R^+)$. In our calculations we use the interpolator $J=\mathcal{P}$, as this gives a better overlap with the ground state than $\mathcal{A}_4$.%
\par%
The renormalized moments in the $\MSbar$ scheme read
\begin{align}
 \langle \xi ^2 \rangle ^{\MSbar} &= \zeta _{11} R^- + \zeta _{12}R^+ ,\\
   a_2 ^{\MSbar} &= \frac{7}{12} \bigl[ 5\zeta _{11} R^- + (5\zeta _{12} - \zeta_{22} )R^+ \bigr],
\end{align}
where $\zeta _{ij}$ are ratios of renormalization constants which are defined
in Ref.~\cite{Braun:2015axa}.

\subsection{Momentum smearing \label{sec_smearing}}
On a lattice of $N_s^3N_t^{}$ sites, separated by the lattice constant
$a$, the linear spatial extent
is given as $L=N_sa$ and spatial momentum components are quantized in terms of
integer multiples of $2\pi/L$. The calculation of the second moment of the
DA requires a spatial momentum~$\vect{p}=(2\pi/L)\vect{n}_{\vect{p}}$,
with at least two non-vanishing components, i.e.,
$\vect{n}_{\vect{p}}^2 \geq 2$. This, in addition to employing two
derivatives, considerably deteriorates the signal-to-noise ratio.
This problem is ameliorated by using momentum smearing~\cite{Bali:2016lva}.
Here we briefly summarize this method.

It is well known that spatially smearing the quark creation and destruction
operators used within the construction of hadronic interpolating fields
increases the overlap of the generated superposition of hadronic states
with the ground state within a given channel. This is not surprising, as
ground state hadrons have smooth, spatially extended wave functions.
The smearing operator~$F$ should be self-adjoint, gauge covariant and
a singlet with respect to all global transformations that act on a
timeslice. In the non-interacting
case its action on a quark field~$q$ can be expressed as a
convolution with a scalar kernel function~$f$:
\begin{align}
\label{smearing}
 (Fq)_{\vect{x}} = \sum _{\vect{y}} f(\vect{x}-\vect{y})q_{\vect{y}}.
\end{align}
In momentum space this convolution becomes a product.

If our smearing kernel is a real Gaussian, then
in momentum space it will remain a Gaussian
centred around $\vect{k}=\vect{0}$. If the hadron carries a
non-vanishing momentum $\vect{p}$, it is natural to assume
that the quark will carry a momentum fraction $\vect{k}=\zeta\vect{p}$.
We remark that there is no obvious relation between $\zeta$
and the longitudinal momentum fraction $x$ of the light-cone wave function.
A Gaussian wave function with width $\sigma$ that is centred about
the momentum $\vect{k}$ acquires a phase:
\begin{align}
\label{Kernel}
f_{(\vect{k})}(\vect{x}-\vect{y})=f_{(\vect{0})}(\vect{0})\exp\left[-\frac{(\vect{x}-\vect{y})^2}{2\sigma^2}+i\vect{k}(\vect{x}-\vect{y})\right],
\end{align}
where $f_{(\vect{0})}=f$. Our periodic lattice appears to imply a quantization of the possible
values of $\vect{k}$. However, Eq.~\eqref{Kernel} can also be cast into
an iterative process, lifting this limitation: It is well known that in
infinite volume the above convolution $F_{(\vect{k})}q$ can be
obtained as the result of evolving the heat equation with a drift term,
\begin{align}
\label{Heat}
\frac{\partial q(\tau)}{\partial\tau}=
\alpha(\vect{\nabla}-i\vect{k})^2 q(\tau),
\end{align}
starting from a spatial delta source at $\tau=0$, to the fictitious time
$\tau=\sigma^2/(2\alpha)$.

One can approximate Eq.~\eqref{Heat} by a
discrete process, defining $F_{(\vect{k})} =\Phi_{(\vect{k})}^n$ as the $n$th
application of an elementary iteration,
\begin{align}
\label{Smear}
(\Phi_{(\vect{k})} q)_{\vect{x}} = \frac{1}{1+6\varepsilon} \left[q_{\vect{x}}+\varepsilon \smashoperator{\sum_{j=\pm 1}^{\pm 3}} U_{\vect{x},j}e^{-i\vect{k}\jhat} q_{\vect{x}+\jhat}\right].
\end{align}
In practice this smearing is implemented by multiplying
the spatial connectors within the timeslice in question by the
appropriate phases, $U_{\vect{x},j}\mapsto e^{-iak_j}U_{\vect{x},j}$.
For $\vect{k}=\vect{0}$ Eq.~\eqref{Smear} corresponds to the
well-known Wuppertal smearing~\cite{Gusken:1989ad,Gusken:1989qx}.
The time coordinate is suppressed as the smearing is local in time.

The gauge connectors within Eq.~\eqref{Smear}, $U_{\vect{x},j}$ and
$U_{\vect{x},-j}\equiv U_{\vect{x}-\jhat,j}^{\dagger}$, where~$\jhat$ denotes
a vector of length~$a$ and direction~$j$, are
spatially APE smeared~\cite{Falcioni:1984ei}:
\begin{align}
\label{APE_Smear}
 U_{\vect{x},i}^{(m+1)} = \mathcal{P}_{\SU 3} \left( \delta\, U^{(m)}_{\vect{x},i} + \smashoperator{\sum _{|j|\neq i}} U_{\vect{x},i}^{(m)}U_{\vect{x}+\jhat,i}^{(m)}U_{\vect{x}+\ihat,j}^{(m)\dagger} \right),
\end{align}
where $i \in \{1,2,3\}$ and $j \in \{\pm1,\pm2,\pm3\}$. The sum is over the four spatial ``staples'' surrounding $U_{\vect{x},i}$, and
$\mathcal{P}_{\SU3}$ is a gauge covariant projector onto the gauge group \SU 3, defined by maximizing $\operatorname{Re}\operatorname{Tr}\{A^\dagger \mathcal{P}_{\SU 3}(A)\}$.
If the APE smeared links are close to unit fields then the width
parameter of the resulting Gaussian is given
as~\cite{Bali:2016lva}\footnote{The root mean squared width
of the resulting Gaussian will correspond to $\sqrt{3}\sigma$
as we have three spatial dimensions.
This will shrink by a factor $1/\sqrt{2}$ if we consider the
squared wave function and since we will smear both quark and antiquark,
the pion interpolator will be wider by a factor $\sqrt{2}$
than the individual quark fields.}
\begin{align}
\label{sigma}
\sigma\approx\sqrt{2na^2}\sqrt{\frac{\varepsilon}{1+6\varepsilon}},
\end{align}
where large values of $\varepsilon$ will allow for smaller
iteration counts~$n$, but the resulting function will be less smooth.

In the meson case the quark creation operator at the source
needs to be smeared with $F_{(\vect{k})}$ and the quark destruction operator
with $F_{(-\vect{k})}$, while for
baryons all three quarks should be smeared with $F_{(\vect{k})}$,
see Ref.~\cite{Bali:2016lva} for details.\footnote{The sign of the complex
phase in Eqs.~\eqref{Kernel}, \eqref{Heat} and~\eqref{Smear} is opposite
to that of Ref.~\cite{Bali:2016lva}. Here we assume that
the phase of the momentum projection at the sink reads
$e^{-i\vect{p}\vect{x}}$ and $\vect{k}=\zeta\vect{p}$ with
$\zeta\geq 0$. The phase used in Ref.~\cite{Bali:2016lva} corresponds
to the non-standard $e^{+i\vect{p}\vect{x}}$ convention that
is used within the {\sc Chroma} software suite~\cite{Edwards:2004sx}.}

\section{Results}
We illustrate the reduction of statistical errors of the two-point functions that enter the calculation of the second moment of the pion DA, using the momentum smearing technique. For this purpose we consider four Coordinated Lattice Simulations (CLS) ensembles, listed in Table~\ref{tab_ensembles}. These range from $N_c\approx 1500$ to $N_c\approx 2800$ configurations, separated by four hybrid Monte Carlo molecular dynamics units. The statistical errors were evaluated using the Bootstrap procedure, with $N_{\text{samples}}=500$, combined with the binning method. For the latter we have observed that a binsize of $N_{\text{bin}}=10$ saturates the statistical error.

The gauge links entering the quark smearing were APE smeared according to~Eq.~\eqref{APE_Smear}, employing $25$ iterations with the parameter $\delta=2.5$. We applied both, the standard Wuppertal smearing~\cite{Gusken:1989ad,Gusken:1989qx} and the novel momentum smearing,
i.e., we implemented Eq.~\eqref{Smear} setting $\vect{k}=\vect{0}$ and $\vect{k}\neq\vect{0}$, respectively, and applied $300$ smearing steps with the smearing parameter $\varepsilon = 0.25$. The root mean squared width of the squared pion interpolator wave function can be estimated using Eq.~\eqref{sigma}. This gives $\sqrt{3}\sigma \approx \sqrt{3}\cdot\unit{0.664}{\femto\meter}\approx\unit{1.14}{\femto\meter}$.

After studying the improvement achieved through momentum smearing, we
attempt a chiral extrapolation of our results. Since we are working at a fixed lattice spacing $a\approx \unit{0.0857}{\femto\metre}$, we cannot as yet perform a continuum limit extrapolation.
\begin{table}[tp]%
\centering%
\caption{\label{table_ensembles}List of the ensembles used in this work. $\beta=3.4$ corresponds to the lattice spacing $a\approx \unit{0.0857}{\femto\metre}$
and $N_c$ denotes the number of analysed configurations. A detailed description of these ensembles can be found in Refs.~\cite{Bruno:2014jqa,Bali:2016umi}.\label{tab_ensembles}}%
\begin{widetable}{\linewidth}{ccccccc}%
  \toprule
  id  &  $N_\text{s}$  &  $N_\text{t}$  & $m_\pi \, [\mega\electronvolt]$ &   $m_K \, [\mega\electronvolt]$ &  $m_\pi L$ & $N_c$\\
  \midrule
  H101  & $32$ & $96$  & $420$ & $420$ & $5.8$ & $2000$\\
  H102  & $32$ & $96$  & $355$ & $440$ & $4.9$ & $1997$\\
  H105  & $32$ & $96$  & $280$ & $465$ & $3.9$ & $2833$\\
  C101  & $48$ & $96$  & $222$ & $474$ & $4.6$ & $1552$\\
  \bottomrule  
\end{widetable}%
\end{table}%

\subsection{Optimizing the smearing and the momentum}
\begin{figure}[t]%
\centering%
\includegraphics[width=\linewidth]{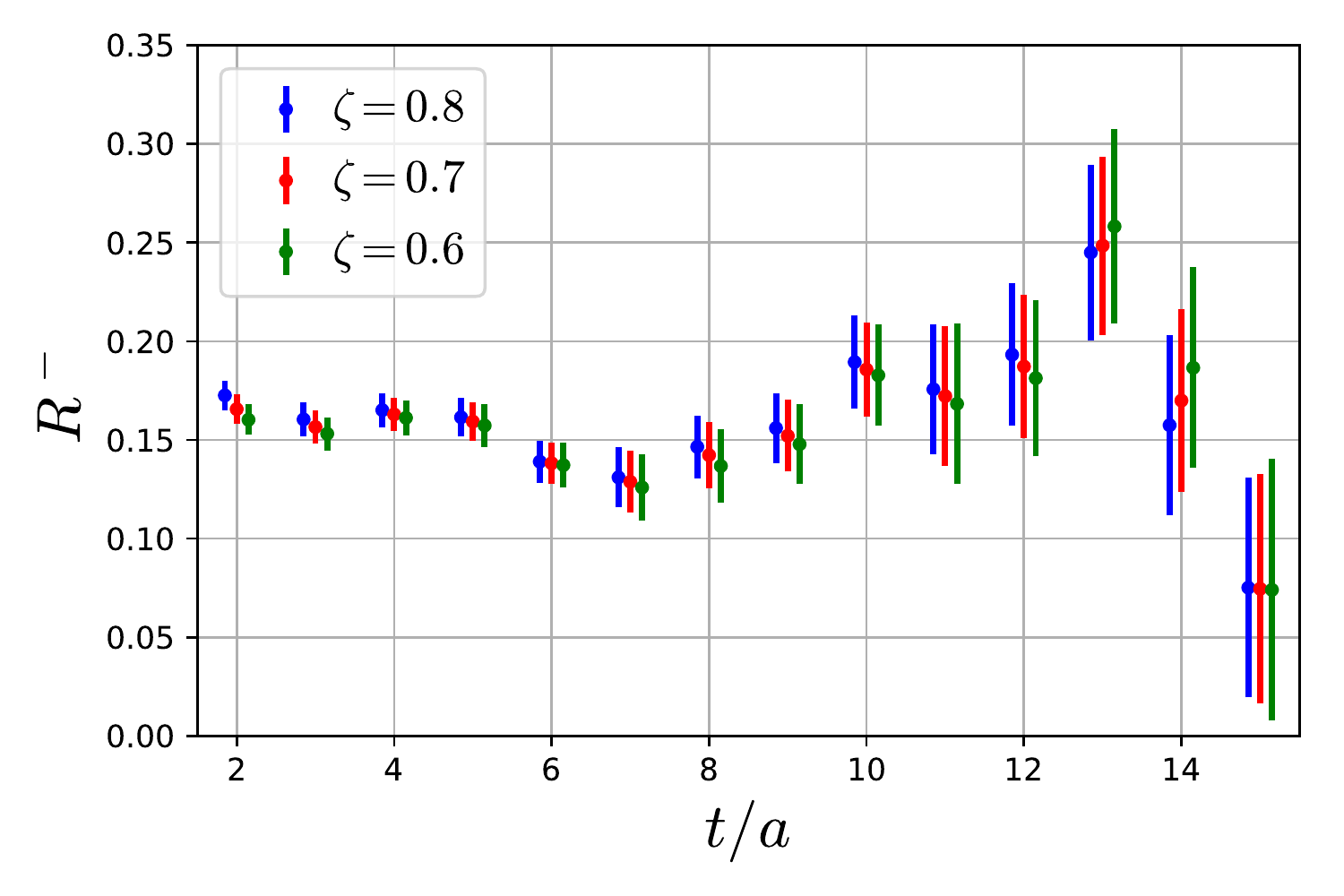}%
\caption{\label{Fig_zetas} Bare lattice value of $R^-$ (see Eq.~\eqref{ratio})
using $331$ configurations of ensemble H105 and the squared momentum $\vect{p}^2=2(2\pi/L)^2 \approx (\unit{0.64}{\giga\electronvolt})^2$.}%
\end{figure}

In order to obtain the second moment of the pion DA we compute ratios of two-point functions that are smeared at the source and local at the sink (smeared-point), where the physical momenta $\vect{p}$ and smearing vectors $\vect{k}$ are parallel:
\begin{align}
 \vect{k}=\zeta \vect{p}.
\end{align}
One may naively expect that a value $\zeta \lesssim 1/2$ was optimal, evenly distributing the meson momentum between quark and antiquark, however, Ref.~\cite{Bali:2016lva} indicated that a value $\zeta\approx 0.8$ was preferable. We confirm this choice: Decreasing $\zeta$ from $0.8$ to $0.6$ we found no improvement of the ground state overlap but slightly increased statistical errors, see Fig.~\ref{Fig_zetas} for the example of the ratio $R^-$, defined in Eq.~\eqref{ratio}. In the following we therefore set $\zeta=0.8$.%
\par%
Using the momentum smearing for mesons, one needs two inversions per momentum vector $\vect{n}_{\vect{p}}$. In contrast to baryonic two-point functions, where all quarks propagate in the forward direction and therefore are smeared using $f_{(\vect{k})}$, the antiquark in mesonic two-point functions needs to be smeared with $f_{(-\vect{k})}$.

\begin{figure}[!t]%
\centering%
\includegraphics[width=\linewidth]{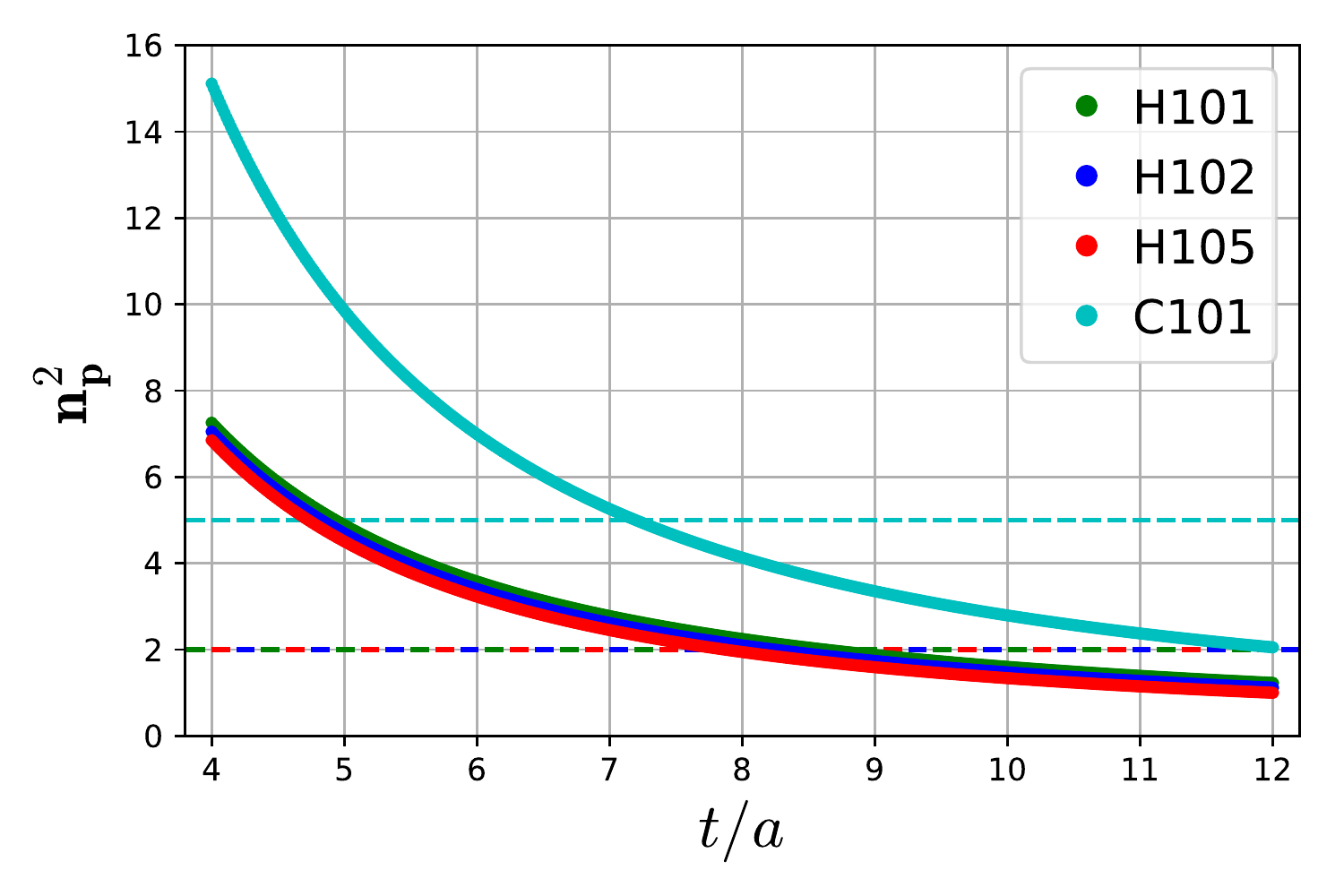}%
\caption{\label{Fig_Gunnar} The optimal $\vect{n}_{\vect{p}}^2$, according to the model Eq.~\eqref{eq_gunnar}, as a function of $t/a$ for each ensemble.}%
\end{figure}

\begin{figure*}[p]%
\centering%
\includegraphics[width=.5\linewidth-.5\columnsep]{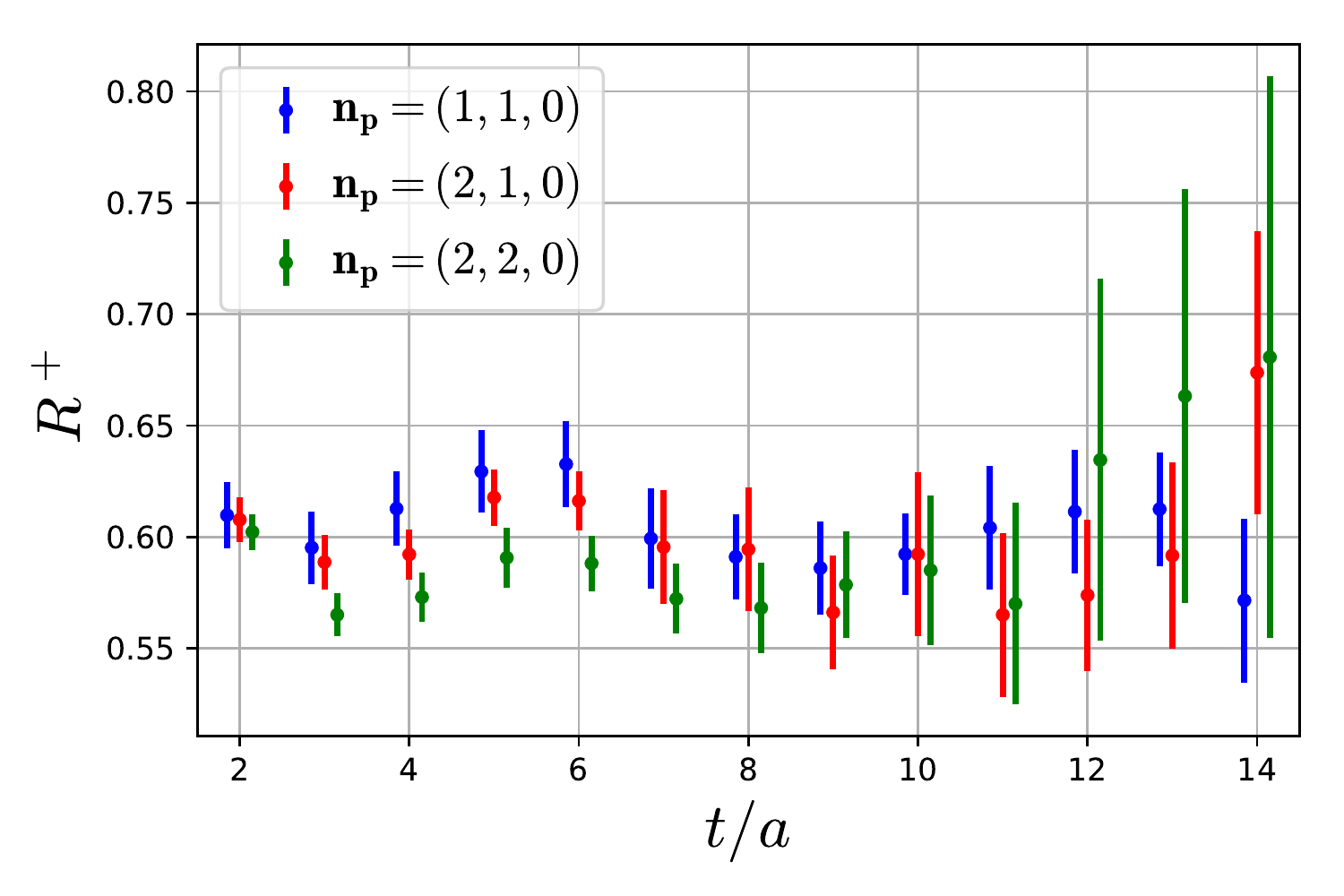}%
\hspace{\columnsep}%
\includegraphics[width=.5\linewidth-.5\columnsep]{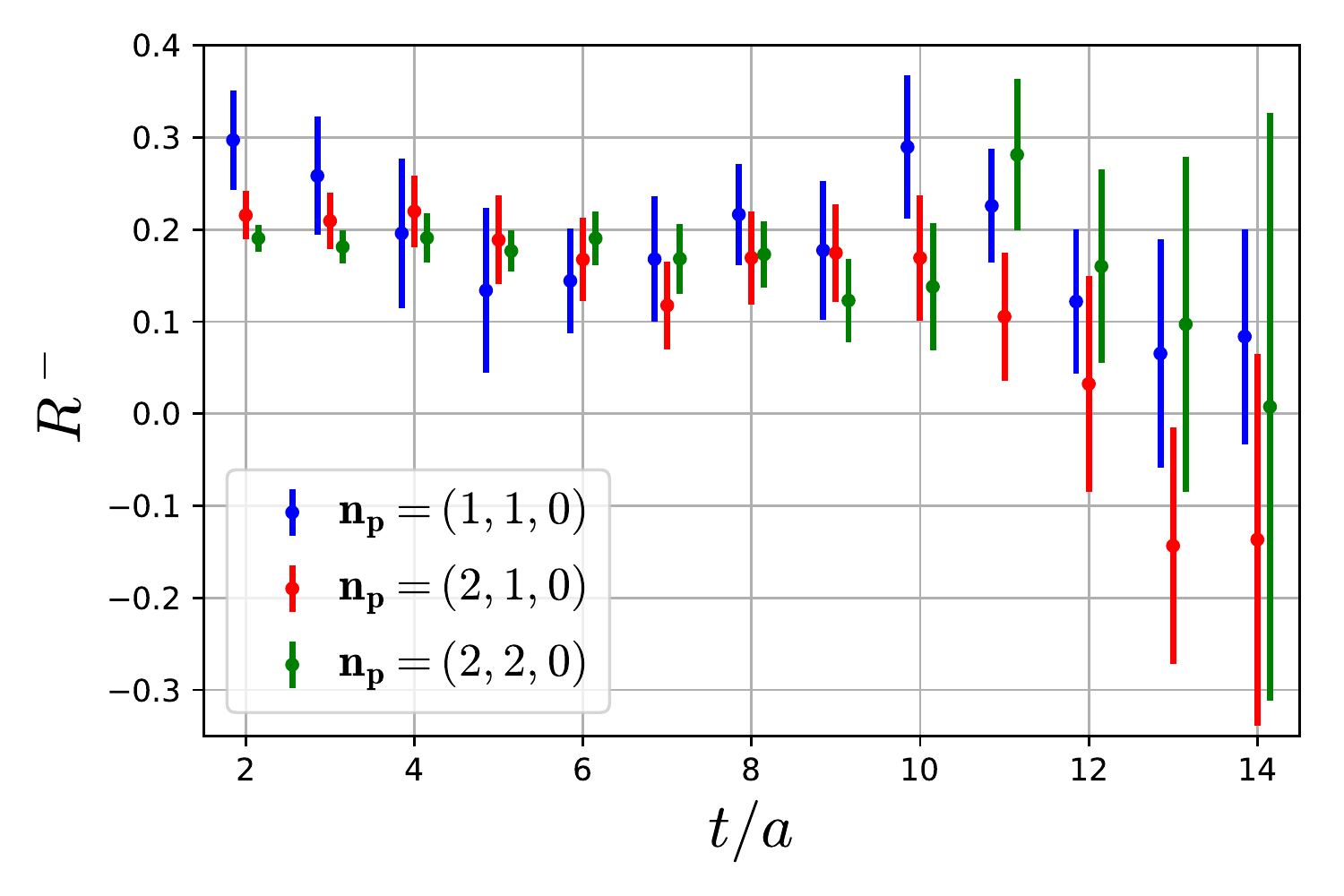}%
\caption{\label{Fig_momenta_C101} Value of $R^\pm$ for C101 with different momenta $\vect{p}$ using $52$ configurations.}%
\end{figure*}

It is instructive to determine which momentum vector~$\vect{n}_{\vect{p}}$  produces the best signal for a given ensemble. Making the crude approximation
of a time-independent noise function, the signal-to-noise ratio of the numerator
that dominates the error of the combination Eq.~\eqref{ratio} can be estimated as
\begin{align}
S(t)\propto p_i p_j \exp\bigl(-\sqrt{m_\pi ^2+\vect{p}^2}t\bigr).
\end{align}
Maximizing this expression with respect to $\vect{p}^2$ gives the positive solution
\begin{align} \label{eq_gunnar}
 \vect{p}^2 = \frac{2}{t^2}\bigl( 1+ \sqrt{1+m_\pi ^2t^2} \bigr).
\end{align}
Clearly, the optimal choice of momentum for a given correlation function depends on $t$ and lower momenta will always be preferred at large values of $t$. This means the outcome will depend on the fit window in $t$ and this in turn will depend on the available statistics.
To aid in finding the most appropriate momentum, we plot Eq.~\eqref{eq_gunnar} in Fig.~\ref{Fig_Gunnar} for the typical fit range for our different ensembles, $t=4a\text{--}12a$. Based on this model, we can read off that for the $L=32a$ lattices squared momenta in the vicinity of $\vect{n}_{\vect{p}}^2=2$ should give reasonable results, whereas for the larger $L=48a$ lattice values of $\vect{n}_{\vect{p}}^2$ closer to $5$ should be investigated. As an example, in Fig.~\ref{Fig_momenta_C101} we show the results of the bare observables $R^\pm$ calculated for different momenta~$\vect{p}$ on the $L=48a$ C101 ensemble. For small values of~$t/a$, larger $\vect{p}^2$ exhibit smaller statistical errors, whereas for large values of $t/a$, the error increases with~$\vect{p}^2$.

In our further analysis we choose $\vect{n}_{\vect{p}}=(1,1,0)$, $\vect{n}_{\vect{p}}=(1,0,1)$ and $\vect{n}_{\vect{p}}=(0,1,1)$ for the ensembles with a spatial extent of $L=32a$ and $\vect{n}_{\vect{p}}=(2,1,0)$, $\vect{n}_{\vect{p}}=(2,0,1)$ and $\vect{n}_{\vect{p}}=(0,2,1)$ for the C101 ensemble. For the $L=32a$ lattices we employ a single source position, while for C101 we realize on average $2$ source positions on each configuration.%
\subsection{Momentum smearing versus Wuppertal smearing}
\begin{figure*}[p]%
\centering%
 \includegraphics[width=.5\linewidth-.5\columnsep]{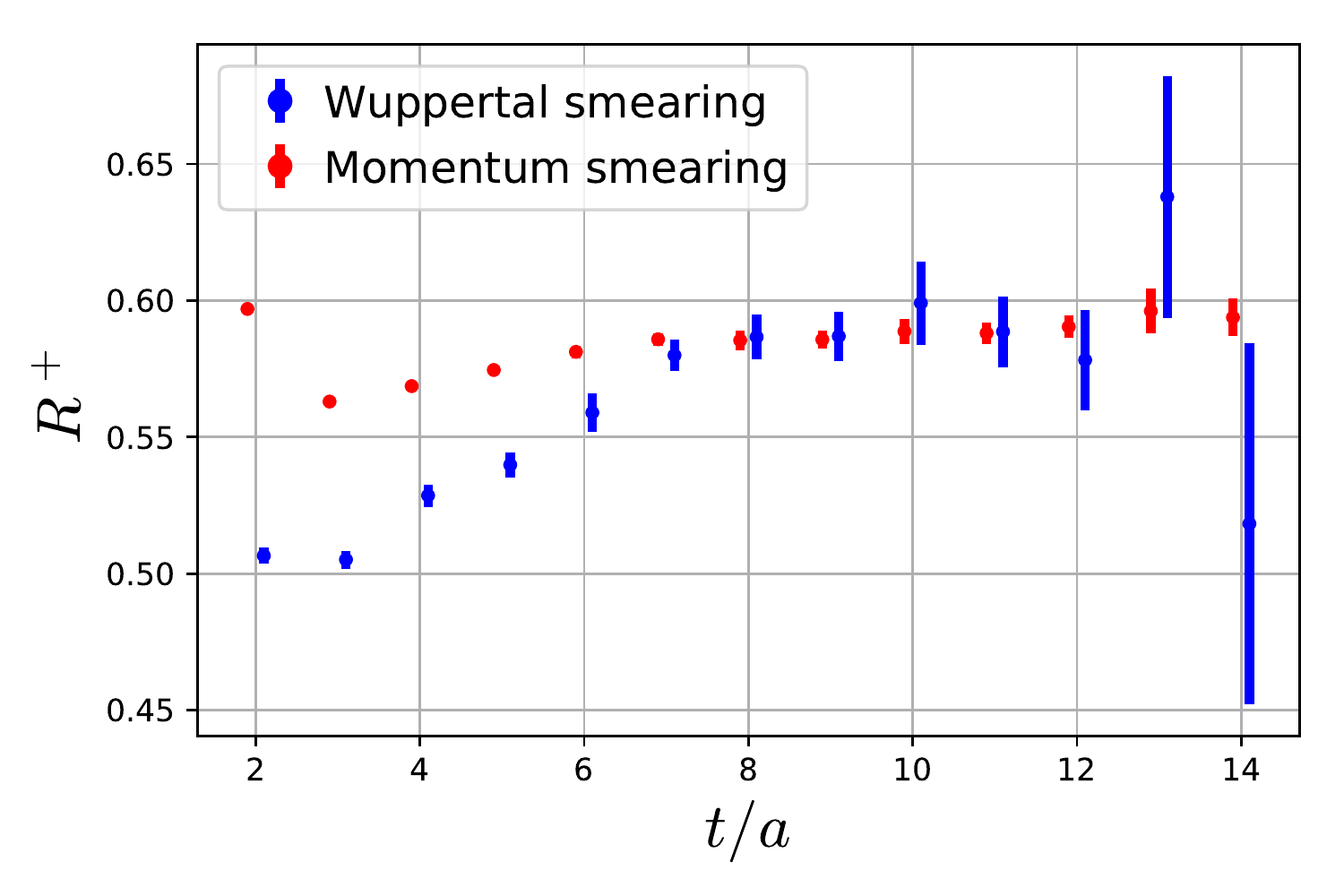}%
 \hspace{\columnsep}%
 \includegraphics[width=.5\linewidth-.5\columnsep]{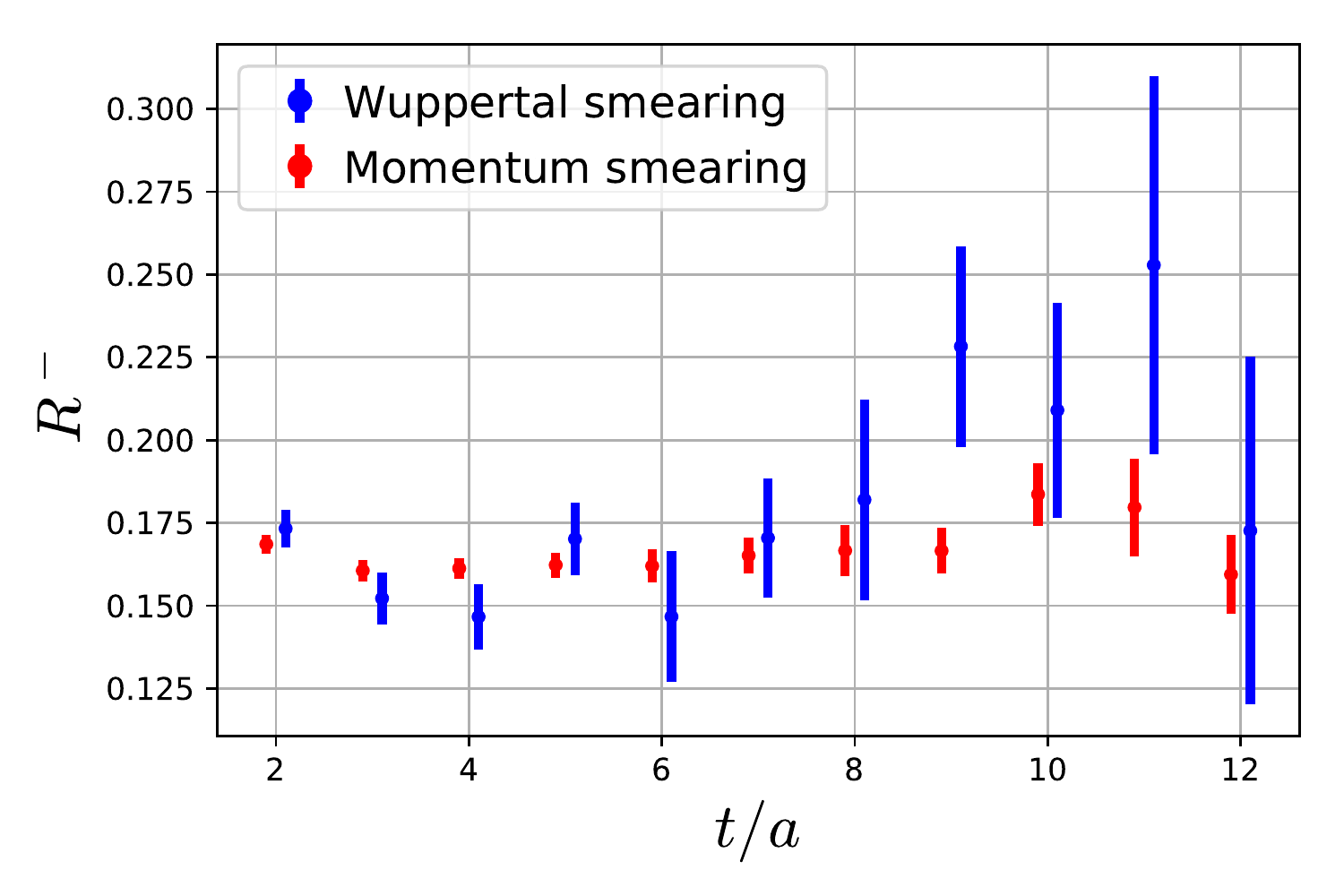}%
\caption{\label{Fig_Plateau} Comparison of the bare lattice values of $R^-$ for H105 using the standard Wuppertal smearing and the new momentum smearing techniques.}%
\end{figure*}

Figure~\ref{Fig_Plateau} shows the plateau of $R^+$ (left) and $R^-$ (right) for the H105 lattice with $N_c = 2830$ for both smearing methods. Clearly the momentum smearing generates a much cleaner and longer plateau with very small statistical errors. In contrast to the standard Wuppertal smearing, where errors increase rapidly for high values of $t$, the signal-to-noise problem is less severe for the momentum smearing. Moreover, in some cases, such as $R^+$ (shown in Fig.~\ref{Fig_Plateau}), we notice reduced contaminations from excited states.

In Fig.~\ref{Fig_Plateau} we compared the results for the same number of configurations. However, the novel momentum smearing method is computationally more expensive. In general one can average over momenta that are equivalent in terms of the cubic symmetry group $\mathrm{O}_h$. Taking into account that the results for $\vect{p}$ and $-\vect{p}$ are trivially related, this gives, depending on the momentum, up to 12 possible lattice directions. For the Wuppertal smearing, additional momenta are computationally almost for free, as they only require additional Fourier sums.
In contrast, for the momentum smearing each momentum direction requires new, differently smeared sources. For the pion two inversions, with momenta $\vect{k}$ and $-\vect{k}$, are necessary as discussed in Sec.~\ref{sec_smearing}. For $\vect{n}_{\vect{p}}^2=2$ this means that momentum smearing is by a factor of almost $6$ more expensive than Wuppertal smearing. Therefore, in Table~\ref{table_cost} we provide an equal cost comparison of the ratios of errors obtained using both methods for the H105 lattice. Even at equal cost, we still see a reduction of the squared error by up to a factor~$3$, in particular for the physically more relevant $R^-$ ratio.
Note that for mesons containing non-degenerate quarks, the traditional method becomes more expensive as this will also require two 
inversions, while for baryon interpolators no momentum smearing with $-\vect{k}$ is required. This means that in terms of a real cost comparison the pion is the least favourable case for momentum smearing.
\begin{table}[tp]%
\centering%
\caption{\label{table_cost} Equal cost comparison of the errors obtained with both methods on the H105 lattice for $R^\pm$. We show the squared ratio of the statistical errors of Wuppertal smearing over momentum smearing divided by the number of inversions needed for each method. }%
\begin{widetable}{\linewidth}{ccccccccccc}%
  \toprule
  $t/a$  &  $3$  &  $4$ &  $5$  &  $6$ &  $7$  &  $8$ &  $9$  &  $10$ &  $11$  &  $12$  \\
  \midrule
  $R^-$  & $1.01$ &  $1.64$ &  $1.35$ & $2.60$ &  $1.82$ &  $2.51$ &  $3.20$ &   $1.96$ &  $2.48$ & $3.21$  \\
  $R^+$  & $0.85$ &  $0.97$ &  $1.03$ & $1.64$ &  $1.03$ &  $0.85$ &  $1.28$ &   $1.76$ &  $1.77$ & $3.24$  \\
  \bottomrule  
\end{widetable}%
\end{table}%

For a fixed number of measurements the gain of momentum smearing is even larger than at a fixed computational cost. However, the reduction of errors that can be achieved by increasing the number of measurements on each configuration is limited, as additional measurements will become increasingly correlated.

\begin{figure*}[p]%
\centering%
\includegraphics[width=.5\linewidth-.5\columnsep]{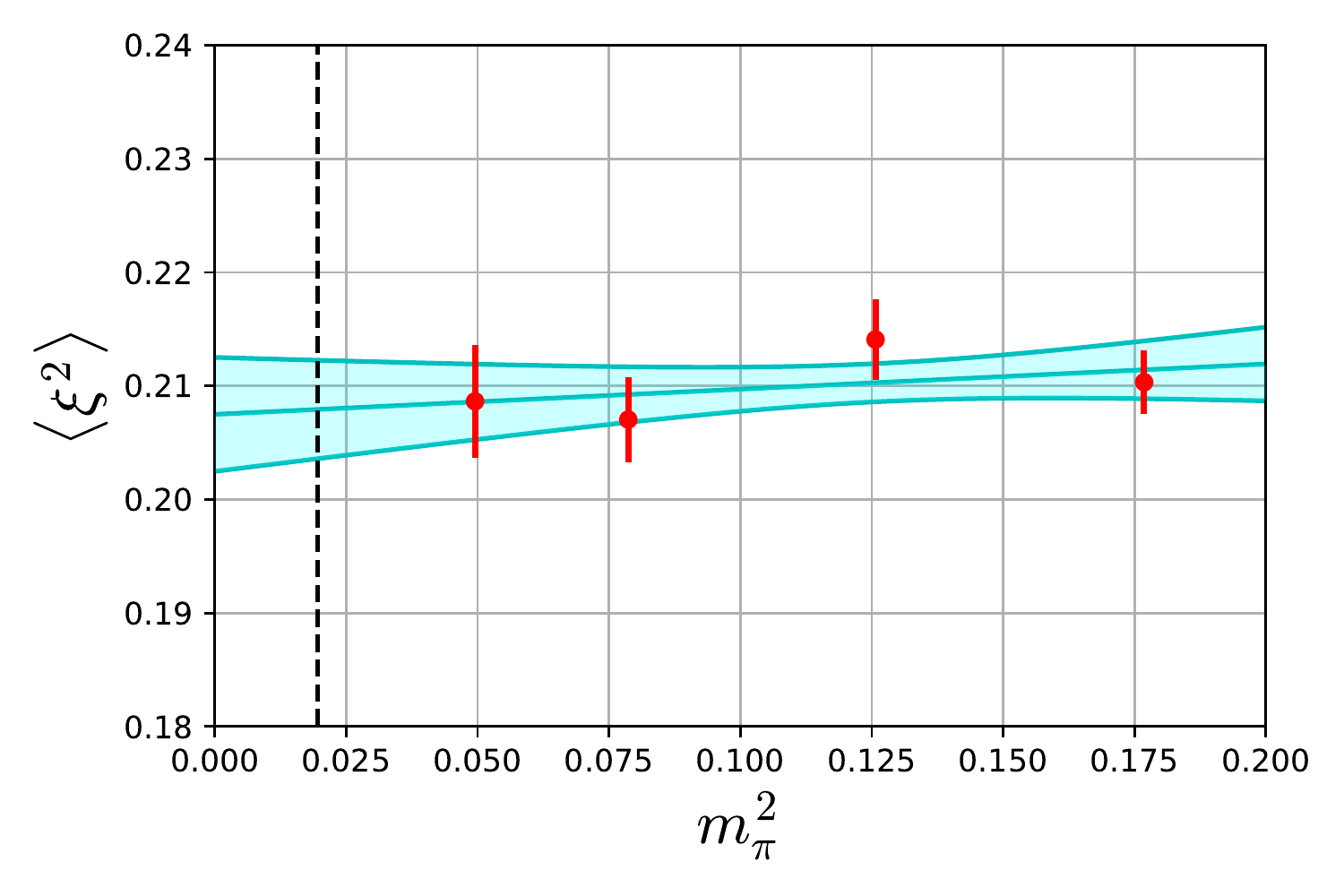}%
\hspace{\columnsep}%
\includegraphics[width=.5\linewidth-.5\columnsep]{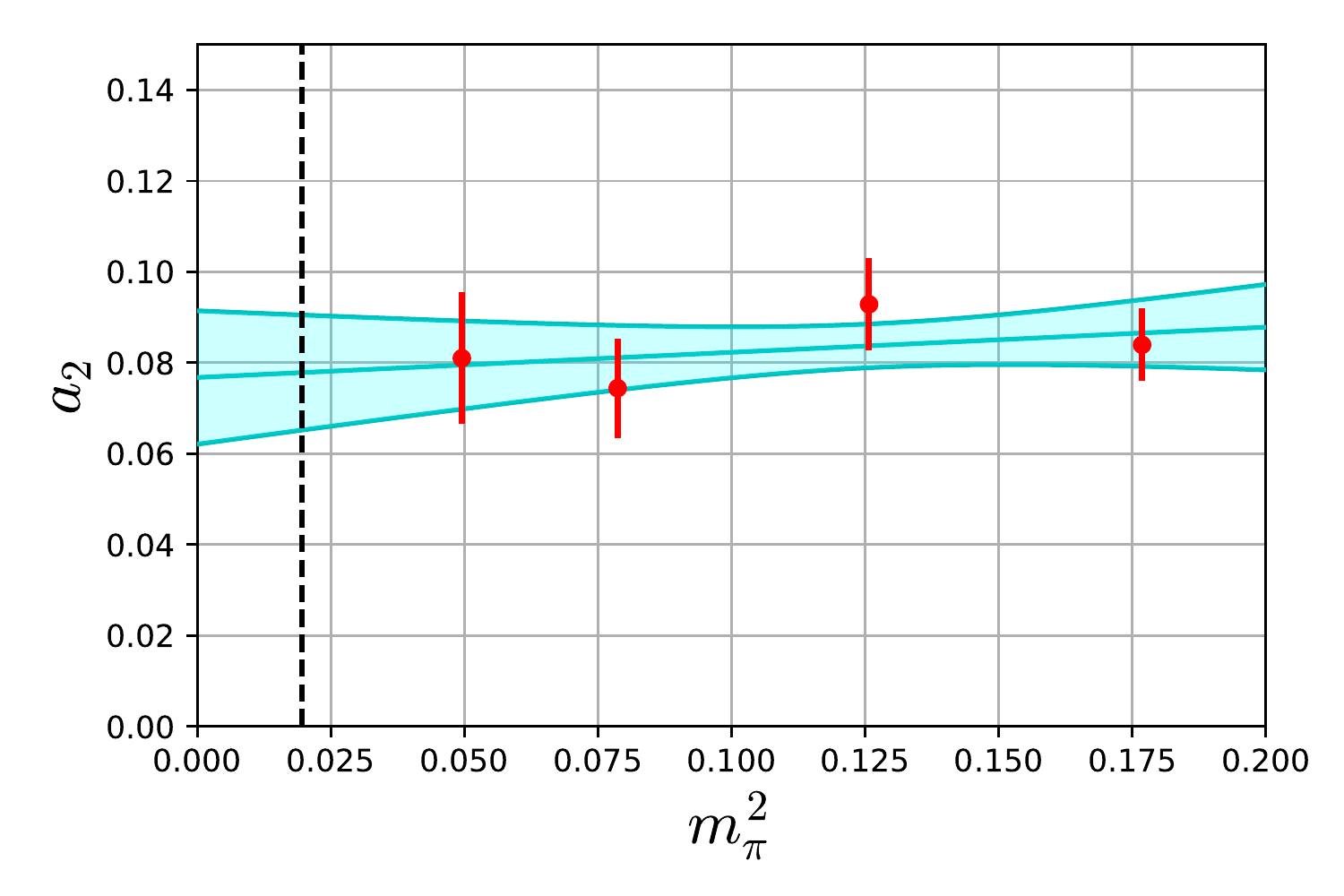}%
\caption{\label{Fig_Extrapolation}Mean and one standard deviation error bands of the chiral extrapolation of $\langle\xi^2 \rangle$ (left) and $a_2$ (right). The vertical dotted line indicates the physical pion mass.}%
\end{figure*}%

\subsection{Chiral extrapolation}
We use Chiral Perturbation Theory (ChPT) to extrapolate our results to physical quark masses. The CLS lattice ensembles used in this work are chosen such that they lie on the $\operatorname{Tr}M=\text{const.}$ line~\cite{Bruno:2014jqa,Bali:2016umi}, which means that to next-to-leading order SU(3)
ChPT the average quadratic meson mass,
\begin{align}
2m_K ^2 + m_\pi ^2,
\end{align}
is kept fixed at its physical value, up to lattice spacing effects. Thus the extrapolation $m_\pi^{\vphantom{\text{phys}}} \to m_\pi ^{\text{phys}}$ also corresponds to $m_K^{\vphantom{\text{phys}}} \to m_K ^{\text{phys}}$.
Up to one-loop order, $\langle \xi^2 \rangle ^{\MSbar}$ and $a_2 ^{\MSbar}$ do not contain chiral logarithms~\cite{Chen:2005js}, and we assume a linear behaviour in $m_\pi ^2$,
\begin{align}
 \langle\xi^2\rangle &= \langle\xi^2\rangle^{(0)} + \langle\xi^2\rangle^{(2)} m_\pi ^2, \\
  a_2  &= a_2 ^{(0)} + a_2 ^{(2)} m_\pi ^2,
\end{align}
where $\langle\xi^2\rangle^{(n)}$ are LECs of the fit. The chiral extrapolation is depicted in Fig.~\ref{Fig_Extrapolation}. At the physical point we find
\begin{align}
 \langle\xi^2\rangle ^{\MSbar} (\unit{2}{\giga\electronvolt}) &=0.2077(43), \\
 a_2^{\MSbar}(\unit{2}{\giga\electronvolt}) &=0.0762(127).
\end{align}
We remark that these numbers were obtained at the fixed lattice spacing $a\approx \unit{0.0857}{\femto\meter}$ and no continuum limit has been performed yet.

\section{Summary}
In this work we have illustrated the effectiveness and advantages of the novel momentum smearing method compared to the standard Wuppertal smearing. For the special case of the pion the momentum smearing requires more inversions, relative to Wuppertal smearing, than for baryons or for mesons consisting of mass non-degenerate quarks. Nevertheless, we have still obtained smaller errors at a similar computational effort. Clearly, using the momentum smearing technique on a fixed number of available configurations, much smaller statistical errors can be achieved. Since for each momentum a new inversion is required in any case, one may suspect that combining the momentum smearing method with the stochastic one-end-trick~\cite{Foster:1998wu} even bigger gains can be achieved. We have not investigated this possibility as yet.

In future studies the momentum smearing will be applied to mesons and baryons on additional CLS lattices, including ensembles at (nearly) physical quark masses and various lattice spacings down to $a \approx \unit{0.04}{\femto\meter}$. This will expand our previous work on meson and baryon DAs~\cite{Braun:2015axa,Bali:2015ykx} and enable us to perform systematic continuum limit extrapolations for mesons and octet baryons.

\section*{Acknowledgements}
This work has been supported by the Deut\-sche For\-schungs\-ge\-mein\-schaft (SFB/TRR\nobreakdash-55) and the Stu\-di\-en\-stif\-tung des deut\-schen Vol\-kes. A significant part of the analysis was carried out on the QPACE~2~\cite{Arts:2015jia} Xeon~Phi installation of the SFB/TRR\nobreakdash-55 in Regensburg. Additional computations were performed on computers of various institutions which we acknowledge below. The ensembles were generated using {\sc openQCD}~\cite{Luscher:2012av}. We used a modified version of the {\sc Chroma}~\cite{Edwards:2004sx} software package along with the {\sc Lib\-Hadron\-Analysis} library and the multigrid solver implementation of Ref.~\cite{Heybrock:2015kpy} (see also Ref.~\cite{Frommer:2013fsa}) to generate hadronic two-point functions. We thank Benjamin Gl\"a\ss{}le and Daniel Richtmann for code development, discussions and software support. Last but not least we thank all our CLS colleagues.
\par
The authors gratefully acknowledge the In\-ter\-dis\-ci\-plin\-ary Centre for Mathematical and Computational
Modelling (ICM) of the University of Warsaw for computer time
on Okeanos (grant No.\ GA67\nobreakdash-12).
We also acknowledge the Gauss Centre for Supercomputing (GCS) for providing computing time for a GCS Large-Scale Project on the GCS share of the supercomputer SuperMUC at Leib\-niz Supercomputing Centre (LRZ, www.lrz.de). GCS is the alliance of the three national supercomputing centres HLRS (Uni\-ver\-si\-t\"at Stutt\-gart), JSC (For\-schungs\-zen\-trum J\"u\-lich), and LRZ (Bay\-e\-ri\-sche Aka\-de\-mie der Wis\-sen\-schaf\-ten), funded by the German Federal Ministry of Education and Research (BMBF) and the German State Ministries for Research of Ba\-den-W{\"u}rt\-tem\-berg (MWK), Bay\-ern (StMWFK) and Nord\-rhein-West\-fa\-len (MIWF).
\par
\bibliographystyle{utcaps}
\bibliography{bibliography.bib}
\end{document}